\documentclass[a4paper,12pt]{article}
\usepackage[pctex32]{graphics}

\begin{document}
\topmargin 0pt
\oddsidemargin 0mm

\newcommand{\alp}{\alpha}
\newcommand{\bta}{\beta}
\newcommand{\gmm}{\gamma}
\newcommand{\del}{\delta}
\newcommand{\omg}{\omega}
\newcommand{\sgm}{\sigma}
\newcommand{\lmd}{\lambda}
\newcommand{\tha}{\theta}
\newcommand{\vph}{\varphi}
\newcommand{\Alp}{\Alpha}
\newcommand{\Bta}{\Beta}
\newcommand{\Gmm}{\Gamma}
\newcommand{\Del}{\Delta}
\newcommand{\Omg}{\Omega}
\newcommand{\Sgm}{\Sigma}
\newcommand{\Lmd}{\Lambda}
\newcommand{\Tha}{\Theta}
\newcommand{\half}{\frac{1}{2}}
\newcommand{\rnd}{\partial}
\newcommand{\nab}{\nabla}

\newcommand{\beqn}{\begin{eqnarray}}
\newcommand{\eeqn}{\end{eqnarray}}
\newcommand{\be}{\begin{equation}}
\newcommand{\ee}{\end{equation}}

\begin{titlepage}
\begin{flushright}
INJE-TP-05-08\\
gr-qc/0511104
\end{flushright}

\vspace{5mm}
\begin{center}
{\Large \bf Entanglement system, Casimir energy and black hole }
\vspace{12mm}

{\large   Yun Soo Myung \footnote{e-mail
 address: ysmyung@physics.inje.ac.kr}}
 \\
\vspace{10mm} {\em  Institute of Mathematical Science, Inje
University, Gimhae 621-749, Korea \\ Institute of Theoretical
Science, University of Oregon, Eugene, OR 97403-5203, USA}

\end{center}

\vspace{5mm} \centerline{{\bf{Abstract}}}
 \vspace{5mm}
We investigate the connection between the entanglement system in
Minkowski spacetime and  the black hole using the scaling
analysis.  Here we show that the entanglement system satisfies the
Bekenstein entropy bound. Even though the entropies of two systems
are the same form, the entanglement energy is different from the
black hole energy. Introducing the Casimir energy of the vacuum
energy fluctuations rather than the entanglement energy, it shows
a feature of the black hole energy. Hence  the Casimir energy is
more close to the black hole than the entanglement energy.
Finally, we find that the  entanglement system
 behaves like the black hole  if
 the gravitational effects are included properly.

\end{titlepage}
\newpage
\renewcommand{\thefootnote}{\arabic{footnote}}
\setcounter{footnote}{0} \setcounter{page}{2}

\section{Introduction}
The fact that the black hole has entropy~\cite{Bek1} and it can
radiate~\cite{Hawk} has led to many debates about its origin of
quantum gravity for decades~\cite{Wald}. Although string theory is
a strong candidate for the quantum gravity, it is so far
incomplete to explain the quantum gravity well. Instead the
holographic principle provides important progress in quantum
gravity.  If a system includes a gravity, guided by the black hole
entropy, the system should obey a suitable entropy
bound~\cite{Hooft,Suss}. The AdS/CFT correspondence  is a
realization of the holographic principle~\cite{Mald}. Also there
were much progress in cosmic holography\cite{FS}.

Usually the local field theory overcounts available degrees of
freedom because it fails to account the effects of gravity
appropriately~\cite{Bous1}. If the gravity is included, not all
degrees of freedom  conjectured by the local field theory can be
used for generating entropy or storing information. This is why we
need to introduce the holographic principle.

On the other hand, the holography could be realized in Minkowski
spacetime.  These are the entanglement system and Casimir effect.
The main features of entanglement system  are summarized by the
observer-dependent entropy, the role of  quantum correlations
across the boundary of the system, and the
non-extensiveness~\cite{BKLS,Sred}. Actually, the entanglement
entropy which is proportional to the area of the boundary is a
measure of an observer's lack of information regarding the quantum
state of the other system in an inaccessible region. In general,
the Casimir effect appears if the system described by the local
(conformal) field theory has the boundary~\cite{Casi}. In this
work we define the Casimir energy by the non-extensive scaling
behavior of $E_C\to \lambda^{1-\frac{2}{n}}E_C$. The Casimir
energy could be found from  the vacuum energy fluctuations of the
local field theory in the entanglement system~\cite{BEFO}. Also
this energy appears as the result of finite-size effects in the
conformal field theory  defined on the Einstein static universe
~\cite{Verl}.  Here the former  means the Casimir energy in the
bulk, while the latter is related to the CFT on the boundary.

In this work we study the connection between the entanglement
system in Minkowski spacetime and  the black hole using the
scaling analysis. We clarify  the distinction between the
entanglement energy and the black hole energy. Introducing the
Casimir energy instead of the entanglement energy, it shows a main
feature of the black hole energy. Finally, we show that the
entanglement system
 behaves like the black hole  if one
 takes  the gravitational effects into account properly.

\section{Entanglement and black hole thermodynamics}
For an  entanglement system, we  consider the three-dimensional
 spherical  volume $V$ and its enclosed boundary $B$ in the flat spacetime.
 We assume that this system has the local (quantum)
 field theory\footnote{In this work we use notations: $\sim$ for a comparison with the IR cutoffs $(R,R_{EH})$ only;
 $\simeq$ for a comparison with both the IR cutoffs $(R,R_{EH})$ and the UV cutoffs ($\Lambda,m_{pl}$).} with
 the IR cutoff $R$  and the UV cutoff $\Lambda=1/a$.
If the vacuum energy density $\rho_{\Lambda}=\Lambda^4$ of the
system does not
 diverge, the vacuum energy and entropy take the forms of $E_{\Lambda} \simeq \Lambda^4R^3$
 and $S_{\Lambda} \simeq \Lambda^3R^3$.

We start by noting the difference between the entanglement
thermodynamics in the flat spacetime and the Schwarzschild black
hole thermodynamics~\cite{MSK}. Here we introduce the first-law of
thermodynamics for  both sides: \be \label{2eq1}
dE_{ENT}=T_{ENT}dS_{ENT},~~ dE_{BH}=T_{BH}dS_{BH}.\ee The
thermodynamic quantities are given by\beqn \label{2eq2}
 S_{ENT} \sim A,~ E_{ENT}\sim A,~ T_{ENT}\sim A^0, \\
\label{2eq3} S_{BH} \sim A_{EH},~ E_{BH}\sim R_{EH},~ T_{BH}\sim
\frac{1}{R_{EH}}, \eeqn where $A=4\pi R^2$ is the
 area of the  boundary $ B $ of the system while
$A_{EH}=4\pi R_{EH}^2$ is the  area of the event horizon.
Eq.(\ref{2eq2}) shows a universal behavior for all entanglement
systems in Minkowski spacetime. Here we note that the zero-point
(vacuum) energy of the system was subtracted in the calculation of
the entanglement energy $E_{ENT}$, thus  degrees of freedom  on
the  boundary contributes to giving  $E_{ENT} \sim A$. It seems
that its areal behavior  is compatible with the concept of the
entanglement. On the black hole side, however, we have  a linear
behavior of the energy. The entanglement entropy behaves
universally as $S_{ENT} \sim A$, which takes the same form as that
of  the black hole. The entanglement temperature is independent of
the radius $R$ of system, whereas the temperature of black hole is
given by $1/R_{EH}$.

 The authors in~\cite{MSK} have explained this discrepancy by noting
that the entanglement quantities are calculated in the flat
spacetime and thus these do not include any gravitational effects.
In order to compare these with those for the black hole, one has
to make corrections to $E_{ENT}$ and $T_{ENT}$ by replacing these
by $E_{ENT}^{new} =\sqrt{-g_{tt}}E_{ENT}$ and
$T_{ENT}^{new}=\sqrt{-g_{tt}}T_{ENT}$ with $\sqrt{-g_{tt}}\simeq
a/R$. It  shows how  an inclusion of gravity alters thermodynamics
of the entanglement system. Considering the connection between $R
\leftrightarrow R_{EH}$ and $a \leftrightarrow l_{pl}$, the new
energy and temperature are the nearly same as those of the black
hole. Here one  observes that  the new entanglement temperature is
red-shifted by inserting the factor of $\sqrt{-g_{tt}}$ at
$r=R+a^2/R$. Simultaneously, the gravitational redshift effect
modifies the area dependence of $E_{ENT}\sim A $ into the linear
dependence of $E^{new}_{ENT} \sim R$, which is the case of   the
black hole. On later, they calculated the entanglement energy in
the Schwarzschild background without introducing   the red-shifted
factor~\cite{MSK1}.

However, the matching procedure between two systems will not end
 at this stage. The above  procedure is not enough  to compare
 the entanglement system with the black hole.
 If one incorporates the gravity to any system,
 the resulting system necessarily
 follow the holographic principle. This is why we need to study
 entropy and energy bounds on matter.
 Also the AdS/CFT correspondence provides a guideline to study the entanglement system when
  including the gravity effects.

\section{Entropy bounds}

In this section we introduce a few of entropy bounds.
 First of all, for a weakly gravitating  system  in
asymptotically flat space, Bekenstein has proposed that the
generalized second law implies the bound~\cite{Bek2} \be
\label{3eq1} S \le S_{B}=\frac{2 k_B \pi E R}{\hbar c}, \ee where
Newton's constant $G$ does not enter here. This bound is powerful
for the system with relatively low density or small volume. For a
black hole with $E_{BH}=Mc^2= c^4R_{BH}/2G$, we find
$S_{BH}=S_{B}$ which means that the Bekenstein bound is saturated
when choosing the black hole as a matter.

 On the other hand, for a
strongly gravitating matter system in the curved spacetime, 't
Hooft and Susskind has shown that the holographic principle
implies the holographic entropy  bound~\cite{Hooft,Suss} \be
\label{3eq2} S \le S_{HOB}=\frac{k_Bc^3A}{4G\hbar}=S_{BH}, \ee
where Newton's constant $G$ is made explicitly.

 Furthermore, Brunstein and Veneziano has
proposed the causal entropy bound~\cite{BV} \be \label{3eq4}
S_{matter}\le S_{CEB}=\frac{c}{2\hbar}\sqrt{\frac{k_BEV}{G}}, \ee
which is given by the geometric mean of $S_{B}$ and $S_{HOB}$.
They showed that for a weakly gravitating
 system\footnote{For the static case, we split all systems which are asymptotically flat
 into  the
weakly self-gravitating system with $R \ge R_{EH}$ and the
strongly self-gravitating system with $R \le R_{EH}$~\cite{Wald}.
For the dynamic case (e.g., cosmology) based on the closed FRW
space of $ds_{FRW}^2=-dt^2+R(t)^2d\Omega_n^2$, we split all
systems into the weakly self-gravitating system with $HR \le 1$
and the strongly self-gravitating system with $HR \ge
1$~\cite{Verl}. Here $H(R)$ are the Hubble parameter (scale
parameter).}, there exists an important sequence \be \label{3eq5}
S_{B}\le S_{CEB} \le S_{HOB}. \ee  This means that for  isolated
systems of weakly gravity, the strongest bound is the Bekenstein
bound while the weakest one is the holographic entropy bound.

In order to carry out the scaling analysis, we use the Planck
units as \be \label{3eq6} \hbar=G=c=k_B=1 \ee in the
$(n+1)$-dimensional spacetimes with $n=3$. Here we need to  check
whether or not Newton's constant $G$ exists for any relations.
This procedure is important because the presence of $G$ represents
an  inclusion of gravity. For this purpose, we study the scaling
behavior of the thermodynamic quantities $S$ and $E$ upon choosing
Eq.(\ref{3eq6})~\cite{Verl}. For example, we have the extensive
scaling behavior for the vacuum energy and entropy: $ E_{\Lambda}
\to \lambda E_{\Lambda},~S_{\Lambda} \to \lambda S_{\Lambda}$
under $V \to \lambda V$.  The Bekenstein entropy is appropriate
for describing a weakly self-gravitating system because $S_{B}$ is
super-extensive: it scales $S_{B} \to \lambda^{1+\frac{1}{n}}$
under $V \to \lambda V, E \to \lambda E$. On the other hand, the
holographic entropy bound is suitable for a strongly
self-gravitating system because $S_{HOB}$ is sub-extensive: it
scales $S_{HOB} \to \lambda^{1-\frac{1}{n}}S_{HOB}$ under $V \to
\lambda V$. The covariant entropy bound is unclear since $S_{CEB}$
is extensive: it scales $S_{CEB} \to \lambda S_{CEB}$ under $V \to
\lambda V, E \to \lambda E$. However,  $S_{CEB}$ includes ``$G$",
which implies that it carries with the effect of gravity.

We often  say that the sub-extensive quantity includes the effects
of gravity, while the super-extensive one does not. Explicitly, a
scaling representation of Eq.(\ref{3eq4}) for a weakly
self-gravitating system is given by\be \label{3eq7}
\lambda^{1+\frac{1}{n}} \le \lambda \le \lambda^{1-\frac{1}{n}}
\ee while for a strongly self-gravitating system\footnote{For
example, in cosmology, they found a sequence $ S_{HEB} \le
S_{BOB}, S_{CEB} \le S_{B}$, where $S_{HEB}=(n-1)HV/4G\hbar$ is
the Hubble entropy to define the Hubble entropy bound of $ S \le
S_{HEB}$~\cite{Verl} and $S_{BCB}=A(B)/4G\hbar$ is the entropy to
define the Bousso's covariant entropy bound of $S[L(B)]\le
S_{BCB}$ on the light sheet $L$ of a surface $B$\cite{Bous2}. Here
the Hubble entropy bound is the strongest one, while the weakest
one is the Bekenstein bound.}, it takes the form
 \be
\label{3eq8} \lambda^{1-\frac{1}{n}} \le \lambda \le
\lambda^{1+\frac{1}{n}}.
 \ee
The above shows clearly that the sub-extensive quantities
represent the strong  gravity, while the super-extensive
quantities denote the weak gravity or ``no gravity". Here we
include the case of ``no gravity" because Bousso derived  the
Bekenstein bound from the generalized covariant entropy bound
(GCEB)~\cite{Bous3}. In his derivation, gravity ($G$) plays a
crucial role. Combining the GCEB involving $1/G\hbar$~\cite{FMW}
with the Einstein equation involving $G$ leads to the Bekenstein
bound in Eq.(\ref{3eq1}) without $G$. Hence this bound can be
tested entirely within the local field theory without any use of
the laws of gravity.

Now let us discuss the non-gravitational systems. The entanglement
entropy $S_{ENT}$ and the Casimir entropy $S_C$ show the areal
behavior \be \label{3eq9} S_{ENT} \sim A \sim
\lambda^{1-\frac{1}{n}},~~S_C\sim A \sim \lambda^{1-\frac{1}{n}}
 \ee
  which are sub-extensive even though they have nothing to do with
  the gravity. Here we assume that  $S_C \sim
E_CR$ for the bulk Casimir entropy.

  In this sense  these  scalings are
  related to the holography induced by non-gravitational
  mechanisms.
  In  case of the entanglement system, requiring the tensor
  product structure of the Hilbert space which is caused by the boundary
  between two regions (like the  event horizon in the black hole)
  leads to the entanglement entropy~\cite{BEY}. That is, the entangled nature of  quantum state
  of the system inside and outside the boundary leads to an areal scaling. On the
  other hand, the Casimir energy of $E_C\sim R$ comes from the  finite-size
  effects in the quantum fluctuations of the local field theory,
  and thus disappears unless the system is  finite~\cite{BFV,GPP}.

 We classify the scaling behaviors of  various entropies into three cases:

\begin{eqnarray}
\label{3eq10}&& \bullet {\rm ~super-extensive} \Longrightarrow
{\rm no~ gravity ~or~weak~ gravity}(S_{B}) \nonumber \\
\label{3eq11}&& \bullet {\rm ~extensive} \Longrightarrow {\rm no
~gravity}(S_{\Lambda}){\rm ~or ~gravity}(S_{CEB}) \nonumber \\
\label{3eq12} && \bullet {\rm ~sub-extensive} \Longrightarrow {\rm
strong~ gravity}(S_{HOB}){\rm ~or~holography}(S_{ENT},S_C).
\nonumber
\end{eqnarray}

Although the entanglement entropy $S_{ENT}$ and the black hole
entropy $S_{BH}$ have the same scaling dimension, they are quite
different because $S_{ENT}$ does not include ``$G$". Now we check
which  entropy bound is suitable for describing the entanglement
system. We note again that the entanglement system is in the flat
spacetime. Hence we could  use the Bekenstein bound for it.
Considering Eq.(\ref{3eq1}), one finds a relation  \be
\label{3eq13} S_{ENT} \sim A \le 2 \pi E_{ENT}R \sim V,\ee which
implies that the entanglement system satisfies the Bekenstein
bound very well. Therefore, the entanglement system is one of
holographic systems in Minkowski spacetime. Also the thermal
radiation in a cavity in the flat spacetime respects this
Bekenstein bound~\cite{Bek3}. However, the Casimir entropy
satisfies the holographic entropy bound of $S_C \le S_{HOB}$.

\section{Energy bounds}

We are in a position to discuss the energy bounds. In this section
we restore ``$G=1/m_{pl}^2$".  First of all, let us check the
scaling dimension of various energies: the black hole energy of
$E_{BH} \simeq m_{pl}^2R$; the Casimir energy of $E_C \sim R$; the
entanglement energy of $E_{ENT}\simeq \Lambda^3R^2 $; the vacuum
energy of $E_{\Lambda} \simeq \Lambda^4R^3$. Under $V \to \lambda
V$, we have the following behavior:

\be \label{4eq1} E_{BH} \sim \lambda^{1-\frac{2}{n}},~~E_C
  \sim \lambda^{1-\frac{2}{n}},~E_{ENT}
  \sim \lambda^{1-\frac{1}{n}},~ E_{\Lambda} \sim \lambda,
 \ee
where all except the vacuum energy are sub-extensive. The black
hole includes obviously the gravitational self-energy. The Casimir
energy\footnote{More precisely, if one introduces the vacuum
energy fluctuations in Ref.\cite{BEFO} instead of $E_C$, one finds
a compact relation of Eq.(\ref{4eq6}).} and entanglement energy
have no  gravitational effects. They become sub-extensive due to
their own properties.

 Cohen {\it et al.} have shown that at  the saturation of
the holographic entropy bound in Eq. (\ref{3eq2})~\cite{CKN}, it
includes many states with $R<R_{EH}$ which corresponds to the
strongly self-gravity condition. Therefore, the energy of most
states will be so large that they will collapse to a black hole
which is larger than the system.  This is hard to be accepted.
Explicitly, the local field theory with $E_{\Lambda}\simeq
\Lambda^4R^3$ and $S_{\Lambda}\simeq \Lambda^3R^3$ is  able to
describe a thermodynamic system at temperature $T$ provided that
$T\le \Lambda$. If $T \gg 1/R$, the energy and entropy will be
those for thermal  radiation: $E_{RAD} \simeq T^4R^3$ and $S_{RAD}
\simeq T^3R^3$. At the saturation of $S_{\Lambda}= S_{HOB}$ with
$T_{SAT}=\Lambda$, one has $T^{S}_{SAT}\simeq
m^{2/3}_{pl}/R^{1/3}$. Considering $R_{EH} \simeq T^{S}_{SAT}R^2$,
one obtains a strongly self-gravity relation of $R_{EH} \simeq
m_{pl}^{2/3}R^{5/3} \gg R$ which is not the case. In order to
eliminate these higher states, they proposed a stronger energy
bound  \be \label{4eq2} E_{\Lambda} \le E_{BH},
 \ee
where the energy bound is saturated if  the system is replaced by
the system-size black hole. At the saturation of $E_{\Lambda}=
E_{BH}$, one finds a relation of $R_{EH} \simeq
m_{pl}^{1/2}R^{3/2}
> R$ with $T^{E}_{SAT} \simeq (m_{pl}/R)^{1/2}$.  In addition, Eq.(\ref{4eq2}) corresponds to
 a more restrictive entropy bound of $S_{\Lambda} \le
A^{3/4}$. Thus this energy bound is stronger than the holographic
entropy bound. However, it will not eliminate all states with
$R<R_{EH}$ lying within $R_{EH}$ completely.

One may accept the weakly self-gravity relation of $R_{EH}\le R$
as a energy bound. If one uses the Planck units, this reduces to
$E_{BH}=M \le R$.  This is called the Schwarzschild
limit~\cite{Hooft} and the gravitational stability
condition~\cite{Bous1,Bous2,Bous3}.  In addition, this corresponds
to the small self-gravitation~\cite{Wald}, the
limited-gravity~\cite{BV,BFV}, and the no-collapse
criterion~\cite{EG,Buni}. We note that this relation includes the
gravity effect of $G$. Hence this relation has nothing to do with
physics in the flat spacetime. One may propose a relation of $E
\le R$ in Minkowski spacetime. However, one should be careful in
using this relation because there does not exist such a linear
behavior of the energy except the Casimir energy and a special
case in~\cite{Suss}. For example, if one uses the saturation
($E=R$, $E$= average total energy of the system) of this bound
$E\le R$ in calculating the entanglement entropy, then one finds
the maximal entropy $S_{MAX}\sim A^{3/4}$ but not the entanglement
entropy $S_{ENT}\sim A$~\cite{Buni}.

It would be better to express the weakly and strongly gravitating
systems in terms of the system energy $E$ and the black hole
energy $E_{BH}$~\cite{Verl,Myung}
\begin{eqnarray}
\label{4eq3}&& \bullet {\rm ~weak~gravity~ condition}: E \le E_{BH} \nonumber \\
\label{4eq4}&& \bullet {\rm ~strong~gravity~ condition}: E \ge
E_{BH}. \nonumber
\end{eqnarray}

We note that $E_{ENT} \sim A > R$. It seems that the entanglement
energy satisfies the strong gravity condition. However, this view
is incorrect because the entanglement system has nothing to do
with the gravity. The holographic nature of its energy  comes from
the concept of entanglement in the flat spacetime. The scaling
relations between the energy and entropy are given by  \be
\label{4eq5}E_{BH}R_{EH} \simeq S_{BH},~E_C R\sim S_C, ~E_{ENT}R
\simeq S_{ENT}^{1+\frac{1}{n-1}},~E_{\Lambda} R \simeq
S_{\Lambda}^{1+\frac{1}{n}}. \ee
 Here it seems that the Casimir system is  closer to the black hole
 than the entanglement system without gravity effect. In order to get the closest case to the black hole,
 we introduce the vacuum energy fluctuations of a free massless scalar field in Minkowski space.
The origin of such energy fluctuations is similar to the
entanglement entropy and energy but the following is
different~\cite{BY}: the trace over degrees of freedom is not
performed on them and quantum expectations values in a pure state
are used for calculation rather than using  statistical averages
over a mixed state ~\cite{MSK}.

 Actually, for the case of $R
 \Lambda> \pi$~\cite{BEFO},  the energy dispersion of  vacuum energy fluctuations
 is given by~\cite{Pad1,Pad2}
 \be \label{4eq6}
 \Delta E_{\Lambda} \simeq  \Lambda^2 R,
\ee where $\Delta E_{\Lambda}$ is defined by \be \label{4eq7}
\Delta E_{\Lambda}\equiv \sqrt{\langle(H_{V}-\langle H_{V}
\rangle)^2\rangle}. \ee Here we choose $E_{C} \simeq \Delta
E_{\Lambda}$ as the Casimir energy for the system in Minkowski
space.
 The weak gravity condition  holds for the Casimir energy
too \be \label{4eq8}
 E_C \le E_{BH},
\ee which means that the Casimir energy by itself cannot be
sufficient to form a system-size black hole.   From
Eq.(\ref{4eq8}), one finds the upper bound on the UV cutoff \be
\label{4eq9} \Lambda \le m_{pl} \ee which may be related to the
gravitational stability condition if one includes a large number
of massless fields~\cite{BEFO}. In addition, we obtain the bound
on the Casimir energy density of $\rho_C \equiv E_{C}/V$
~\cite{CKN}\be \label{4eq10} \rho_C
 \simeq \frac{\Lambda^2}{R^2} \le \rho_{BH}\simeq \frac{m_{pl}^2}{R^2}.
\ee On the other hand, there exists the vacuum energy bound from
Eq.(\ref{4eq2}) as \be \label{4eq11} \rho_{\Lambda}\equiv
\Lambda^4 \le \rho_{BH}. \ee This implies that  the vacuum energy
by itself cannot be sufficient to form a system-size black hole.
At the saturation of these bounds, we obtain the important
holographic energy density $\rho_{HOG}=m_{pl}^2/R^2$, which was
used widely to explain the dark energy in
cosmology~\cite{Hsu,Li,Myung}. A relation of
$\rho_{HOL}=\rho_{BH}$ implies that the whole universe is
dominated by black hole states. Thus it is called the maximal
darkness conjecture~\cite{KLL}.

\section{Where is the discrepancy?}
Up to now we do not introduce any effect of gravity in the
entanglement system. It was shown that the entanglement system
behaves like the black hole  if the gravitational effects are
included properly~\cite{MSK}. We wish to reconcile our picture
with the AdS/CFT correspondence, where the gravitational
blue/redshift play the important role in establishing the
holographic principle. For this purpose, we introduce the
($n+2$)-dimensional AdS-black hole with the metric
element~\cite{Witten} \be \label{5eq1}
 ds_{AdS}^2=
-\Big[1+\frac{r^2}{l^2}-\frac{\omega_n M}{r^{n-1}}\Big]dt^2
+\frac{dr^2}{1+\frac{r^2}{l^2}-\frac{\omega_n
M}{r^{n-1}}}+r^2d\Omega_{n} \ee where $l$ is the AdS-curvature
radius related to the cosmological constant
$\Lambda_{n+2}=-n(n+1)/2l^2$ and $\omega_n=16 \pi G_{n+2}/nV_n$
with $G_{n+2}$ the ($n+2$)-dimensional Newton constant and $V_n$
the volume of unit $S^n$. Here the bulk thermodynamic quantities
of black hole energy, Hawking temperature and Bekenstein-Hawking
entropy are given by \be \label{5eq2} E=M=
\frac{r_+^{n-1}}{\omega_n}\Big[\frac{r_+^2}{l^2}+1\Big],~~
T_{H}=\frac{1}{4 \pi}\Big[
\frac{(n+1)r_+}{l^2}+\frac{(n-1)}{r_+}\Big],
~~S_{BH}=\frac{V_n}{4G_{n+2}}r_+^n. \ee According to
Ref.~\cite{Verl}, the Casimir energy is defined by the violation
of the Euler identity as \be \label{Cas1}
E^c_{b}=n(E-T_HS_{BH}+pV). \ee Noting that the CFT is a
radiation-like matter at high temperature, one has the equation of
state for $pV=E/n$. Then the Casimir energy in the bulk is given
by \be\label{Cas2}E^c_{b} =
(n+1)E-nT_HS_{BH}=\frac{nV_nr_+^{n-1}}{8\pi G_{n+2}}. \ee On the
other hand, the ($n+1$)-dimensional CFT near infinity is defined
by the Einstein static universe~\cite{Verl}: \be \label{5eq3}
ds_{ESU}^2=-d\tau^2+\rho^2d\Omega_n^2.\ee Now we can determine the
boundary metric of Eq.(\ref{5eq3}) by using  the bulk metric in
Eq.(\ref{5eq1}) near infinity as
 \be
\label{5eq4} ds_b^2=\lim_{r\to \infty}\frac{\rho^2}{r^2}ds_{AdS}^2
=-d\tau^2+\rho^2d\Omega_n^2 \to ds_{ESU}^2,~~
\tau=\frac{\rho}{l}t. \ee Using the Euclidean formalism, we find
bulk-boundary relations~\cite{CMN}: \be \label{5eq5}
T_{CFT}=\frac{l}{\rho} T_{H}=\frac{1}{4\pi
\rho}\Big[(n+1)\hat{r}+\frac{(n-1)}{\hat{r}}\Big],~
E_{CFT}=\frac{l}{\rho} E=\frac{nV_n \kappa
\hat{r}^{n-1}}{\rho}\Big[\hat{r}^2+1\Big]\ee and \be
\label{cftther} E^c_{CFT}=\frac{l}{\rho}E^c_{b}=\frac{2nV_n\kappa
\hat{r}^{n-1}}{\rho}, ~S_{CFT}=S_{BH}=4\pi \kappa V_n\hat{r}^n \ee
with $\hat{r}=r_+/l$ and  $\kappa=l^n/16\pi G_{n+2}$. Here we can
choose the radius $\rho$ of $S^n$ as we wish. Thus, considering
the duality of $r_+ \leftrightarrow \hat{r}$, we find the same
forms for the CFT quantities as those in the bulk-AdS space. We
note the different functional forms for a large black hole with
$r_+ \gg l$: $E \sim r_+^{n+1},~S_{BH} \sim r_+^n,~E^c_b \sim
r_+^{n-1}$. Also we have the same relations for the dual CFT. This
means that in the AdS spacetime and its dual CFT, the Casimir
energy has the lowest power in compared with the energy and
entropy. As an example, we have $E^c_b \sim r_+$ for
(1+3)-dimensional AdS spacetime and $E^c_{CFT} \sim \hat{r}$ for
(1+2)-dimensional CFT.

 We call these as
either the UV/IR scaling transformation in the AdS/CFT
correspondence or the Tolman redshift transformation on the
gravity side~\cite{GPP2}. The scaling factor of
$\sqrt{-g^{tt}_{\infty}}=l/\rho$ comes from the fact the Killing
vector $\partial/\partial t$ is normalized so that near infinity
\be \label{5eq6} g\Big(\frac{\partial}{\partial
t},\frac{\partial}{\partial t}\Big) \to -\frac{\rho^2}{l^2}.\ee
 This fixes
the  red-shifted CFT of $E,~E^c$, and $T$, but $S$ is not scaled
under the UV/IR transformation. Although there is no room to
accommodate the AdS/CFT correspondence in the entanglement system,
we use it  to point out the discrepancy between the entanglement
and black hole systems.

Now let us introduce the Tolman redshift transformation on the
black hole system. In general, the local temperature observed by
an  observer at $r>R_{EH}$ in the Schwarzschild black hole
background is defined by~\cite{York} \be \label{5eq7} T(r)=
\frac{T_{\infty}}{\sqrt{-g_{tt}}}=\frac{1}{8\pi
M}\frac{1}{\sqrt{1-\frac{2M}{r}}} \ee where $T_\infty=1/8\pi M$ is
the Hawking temperature measured at infinity and the denominator
of $\sqrt{-g_{tt}}$ is the red-shifted factor. Near the horizon
(at $r=R_{EH}+l_{pl}^2/R_{EH}$), this local temperature is given
by $T=1/8 \pi l_{pl}$ which is independent of the black hole mass
$M$. Similarly,  the local energy is given by $E(r)=
\frac{E_{\infty}}{\sqrt{-g_{tt}}}$ with $ E_{\infty}=M=R_{EH}/2$.
Near the horizon, we have $E \propto A_{EH}$. However, there is no
difference between the local entropy $S_{BH}$ near the horizon and
the entropy $S^{\infty}$ at infinity.

The Tolman redshift transformation is also possible between
$E_{ENT},~T_{ENT},~S_{ENT}$ near the boundary $B$ and
$E^{\infty}_{ENT},~T^{\infty}_{ENT},~S^{\infty}_{ENT}$ at infinity
if the gravity effect is included in the entanglement system.
Choosing the local temperature as $T_{ENT}\simeq 1/a$ at
$r=R+a^2/R$, then the corresponding temperature at infinity is
given by $T_{ENT}^{\infty}= \sqrt{-g_{tt}}T_{ENT} \simeq 1/R$
because of $\sqrt{-g_{tt}} \simeq  a/R$ near the horizon.
Furthermore, if the local energy is given by $E_{ENT} \simeq
A/a^3$ at $r=R+a^2/R$, then the corresponding energy at infinity
is $E_{ENT}^{\infty}=\sqrt{-g_{tt}} E_{ENT} \simeq R/a^2$.  The
entanglement entropy remains unchanged under the transformation.
This picture is consistent with Ref.\cite{MSK}. That is, the
entanglement system including the gravity effects shows the
feature of the black hole.

However, there exists a discrepancy in the Casimir energy.
Considering $E_C \sim R/a^2$ as the local Casimir energy, the
Casimir energy at infinity
 is given by $E^{\infty}_C=\sqrt{-g_{tt}}E_C \simeq 1/a $.
Now our question is whether or not Eq.(\ref{Cas2}) can apply to
the asymptotically flat black hole in the $(2+n)$-dimensional
spacetime: $E^{\infty}_S=nV_nr_+^{n-1}/16 \pi
G_{n+2},~S=V_nr_+^n/4G_{n+2}\sim A,~T^{\infty}_S=(n-1)/4\pi r_+$.
These are obtained from Eq.(\ref{5eq2}) together with $r_+ \ll l$.
Let us assume that like the AdS-black hole, this black hole may be
described by a radiation-like  CFT in $(1+n)$
dimensions~\cite{KPSZ}. Then, using Eq.(\ref{Cas2}), we have \be
\label{bcas} \tilde{E}^{\infty}_C= \frac{n V_n}{8 \pi
G_{n+2}}r_+^{n-1}=2E^{\infty}_S, \ee for the Schwarzschild black
hole. This means that the Casimir energy is nearly the same form
as in the black hole energy. Here we conjecture that the local
Casimir energy is given by $\tilde{E}_C \sim A$. Then  we have
$\tilde{E}^{\infty}_C =\sqrt{-g_{tt}}\tilde{E}_C \sim R$ like the
black hole energy in the  $(1+3)$-dimensional spacetime. The
discrepancy between $E^{\infty}_C$ and $\tilde{E}^{\infty}_C$
mainly arises from the handicap of the Schwarzschild black hole
which is defined in asymptotically flat space. Actually
Eq.(\ref{Cas2}) holds only if there exists its dual CFT. However,
we do not know what is its dual CFT. As a result, we cannot
establish the correct connection between the local Casimir energy
and Casimir energy at infinity at this stage. This is mainly
because the Schwarzschild black hole is too simple to split the
energy into the black hole energy and Casimir energy, in contrast
with the AdS-black hole.

\section{Discussions}
We have the two kinds of the  holography. The first one is called
the g-holography induced by the gravity which appeared in  the
black hole and cosmology. The second holography  is shown  by
non-gravitational mechanisms in the flat spacetime.  The
sub-extensive entropy including the gravity effect ``$G$" belongs
to the g-holography. For example, these are $S_{BH}=S_{HOB} $ for
the black hole and  $S_{CEB}, S_{HEB}$ and $ S_{BOB}$ for
cosmology. On the other hand, the sub-extensive  entropy defined
without $G$ can be used to describe the holography realized in the
flat spacetime. There are $S_{ENT}$ and $ S_C$.

Now we discuss the connection between the entanglement system and
black hole. There exists  difference between two systems. The
entanglement energy shows an areal behavior in contrast to the
linear behavior of the black hole. The entanglement system
satisfies the Bekenstein entropy bound which is suitable for
either  the no-gravity  system or the weakly gravitational system.
Also we find that the Casimir energy (vacuum energy fluctuations)
is more close to the black hole energy than the entanglement
energy.

 However, two systems behave like the same if the gravity effects
 are included  in the entanglement system. This is
checked with  the UV/IR scaling transformation in the AdS/CFT
correspondence and the Tolman redshift transformation on the
gravity system.

\section*{Acknowledgment}
The author  thanks Steve Hsu,  Roman Buniy, and Brian Murray for
helpful discussions. This work  was supported by the Korea
Research Foundation Grant (KRF-2005-013-C00018).

\end{document}